\documentclass[12pt]{iopart}
\usepackage{graphicx}

\def\USIR{unstructured {}}
\begin{document}

\title[Propagation and mitigation of epidemics in a scale-free network]{Propagation and mitigation of epidemics in a scale-free network}

\author{Gyula M. Szab\'o}

\address{ELTE E\"otv\"os Lor\'and University, Gothard Astrophysical Observatory and Multidisciplinary Research Center, Szombathely, Hungary}
\ead{szgy@gothard.hu}
\vspace{10pt}
\begin{indented}
\item[]March 2020
\end{indented}

\begin{abstract}
The epidemic curve and the final extent of the COVID-19 pandemic are usually predicted from the rate of early exponential raising using the SIR model. These predictions implicitly assume a full social mixing, which is not plausible generally. Here I am showing a counterexample to the these predictions, based on random propagation of an epidemic in Barab\'asi--Albert scale-free network models. The start of the epidemic suggests $R_0=2.6$, but unlike $\Omega\approx 70\%{}$ predicted by the SIR model, they reach a final extent of only $\Omega\approx 4\%{}$ without external mitigation. Daily infection rate at the top of the curve is also an order of magnitude less than in SIR models. Quarantining only the 1.5\%{} most active superspreaders has similar effect on extent and top infection rate as blind quarantining a random 50\%{} of the full community.
\end{abstract}

\section{Introduction}

The SIR epidemic model (\cite{km1927, lipsitch+2003}) is often invoked to predict the basic pandemic parameters and predictions from the early growth rate. The model assumes a homogeneuous environment and full social mixing. The implicit assumptions are: 1) the expectation number of new infections caused by an ill person is constant in the entire population -- this number is a model parameter, $R_0$, and 2) anyone in the society can be infected by anyone else with the same probability. The persons are initially susceptible (S), after they get the disease and become ill (I), and they infect new persons. Among these, the susceptible persons develop disease and spread again, and those who have been recovered (R) and have immunity do not get ill again and do not spread the disease again. The resulting differential equations predict an exponential initial phase, where the trend reflects $R_0$. If $R_0>1$, the epidemics develop, and if $R_0<1$, it decays out.

When the pandemic spreads, the number of new cases exponentially increase in early stages. Later it flattens and starts decreasing, following a more or less symmetrical hump. The expected duration of the pandemics, $T_p$ and the final extent of the epidemics, $\Omega$ are key parameters in planning all actions to slow down and/or cure the pandemics. The best possible prediction of these parameters even in the early stages of the pandemics is crucial. $\Omega$ can be esimated according to the rule of thumb that $(1-\Omega)R_0<1$, i.e. the case when the number of new infections is less at the next step than in the previous one. 
Current estimates of the COVID-19 disease suggest $R_0=2.2$--$2.6$, indicating that 60--70\%{} of the population will be rapidly infected before the pandemics cease, and the duration of the hump will be roughly 3 months in the UK (\cite{ferguson+2020}).

These predictions quantitatively much rely on the assumed homogeneity of the society and on the full random mixing during the propagation of the disease. Here I am showing a simple example where infections are transmitted via a scale-free social network, and in this case the propagation scenarios can follow a much milder evolution after the early steep upslope than in the homogeneous model. This can base a less pessimistic scenario of the prognosis of the epidemic develompent, despite of its intensive start.

\section{Propagation models}

\begin{figure}\centering
\includegraphics[bb=110 110 400 400,width=5cm,clip]{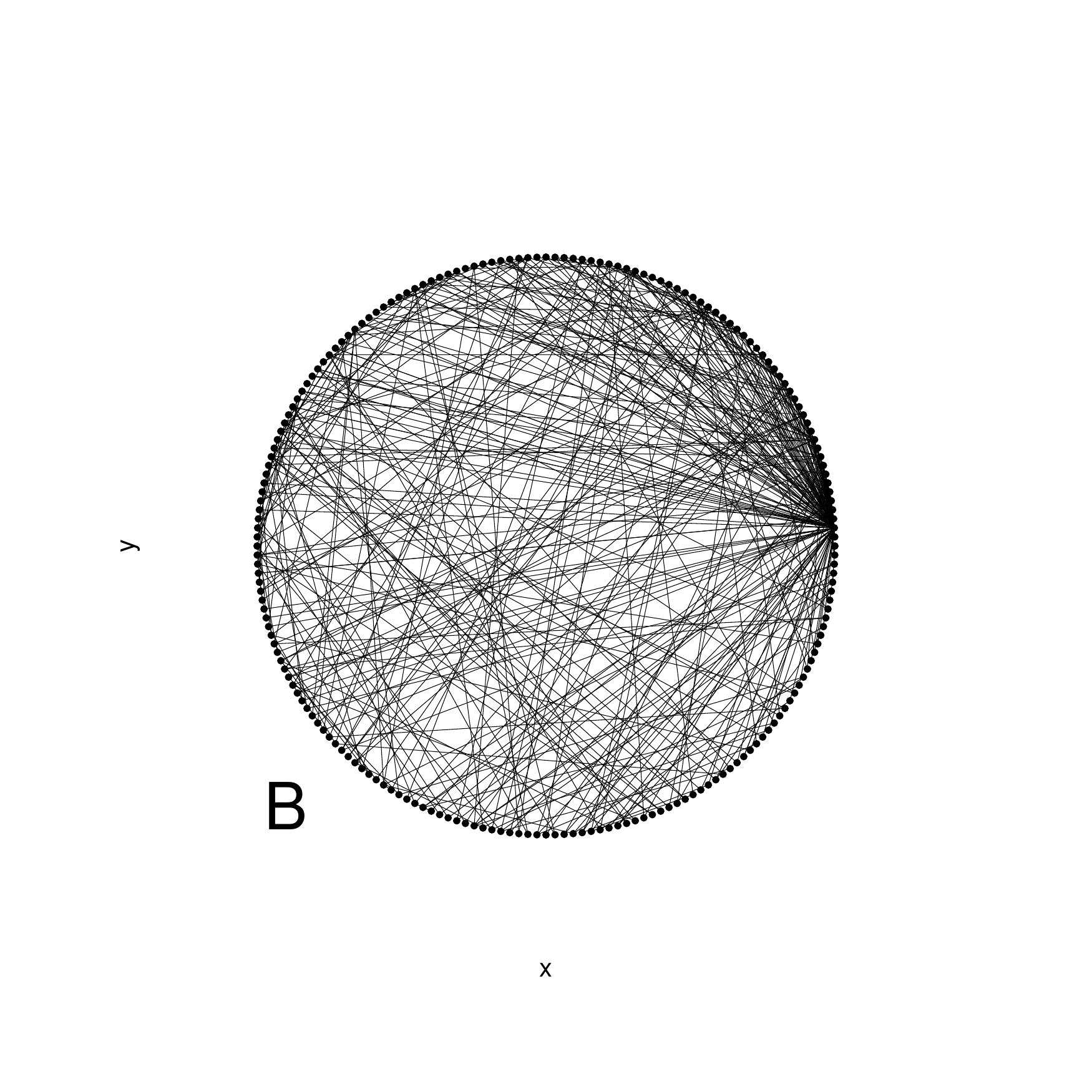}
\includegraphics[bb=110 110 400 400,width=5cm,clip]{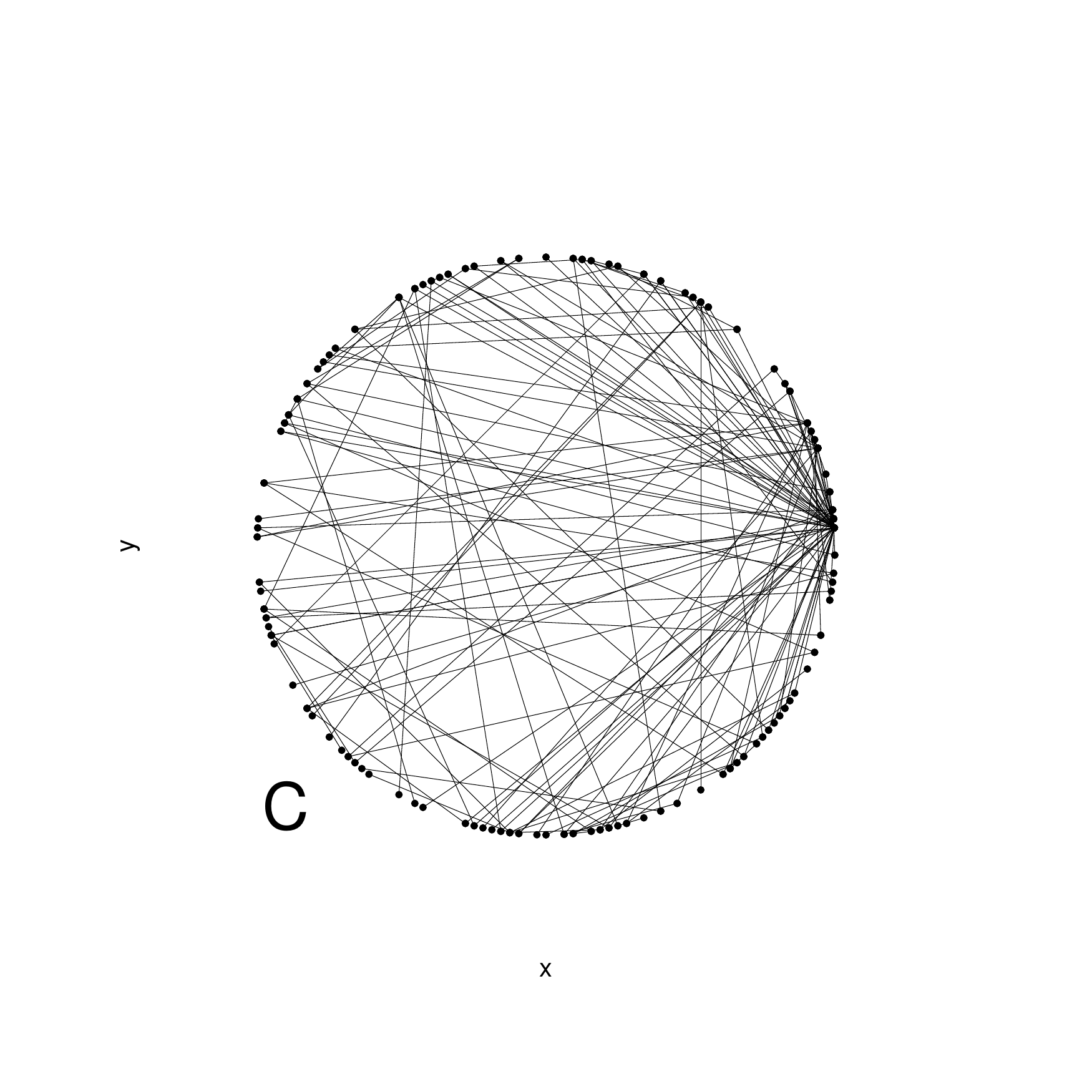}
\includegraphics[bb=110 110 400 400,width=5cm,clip]{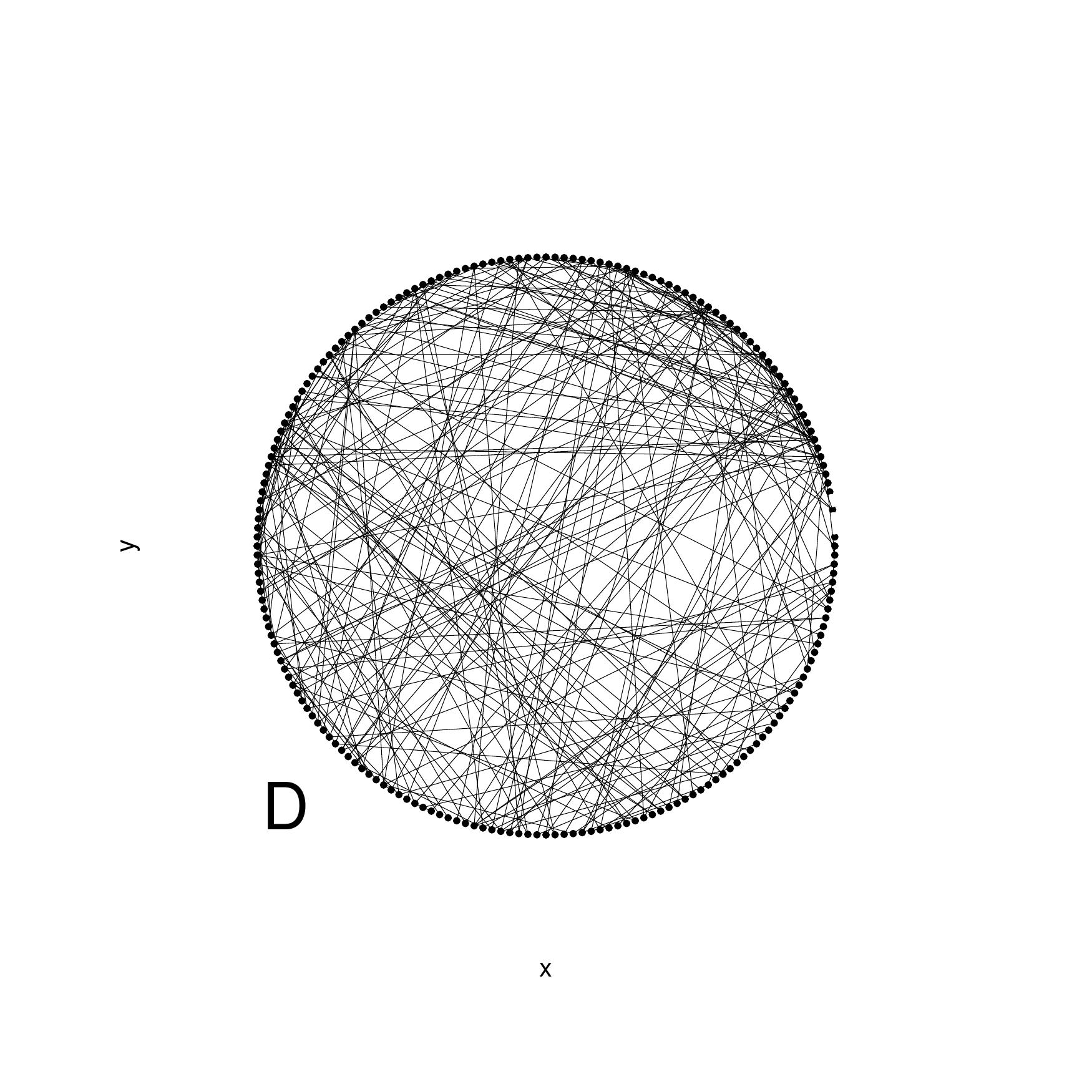}
\caption{Illustration of the networks, but including only 200 persons instead of 50,000 for perspicuity. B: A Barab\'asi--Albert model with $m_0=m=2$, C: Same as B, half of the persons were quarantained; D: Same as B, only the three most active superspreaders quarantained. The \USIR model (labeled as A in the forthcomings) has no graphic representation}\label{fig:networks}
\end{figure}

\begin{figure}\centering
\includegraphics[bb=1 70 440 370,width=8cm,clip]{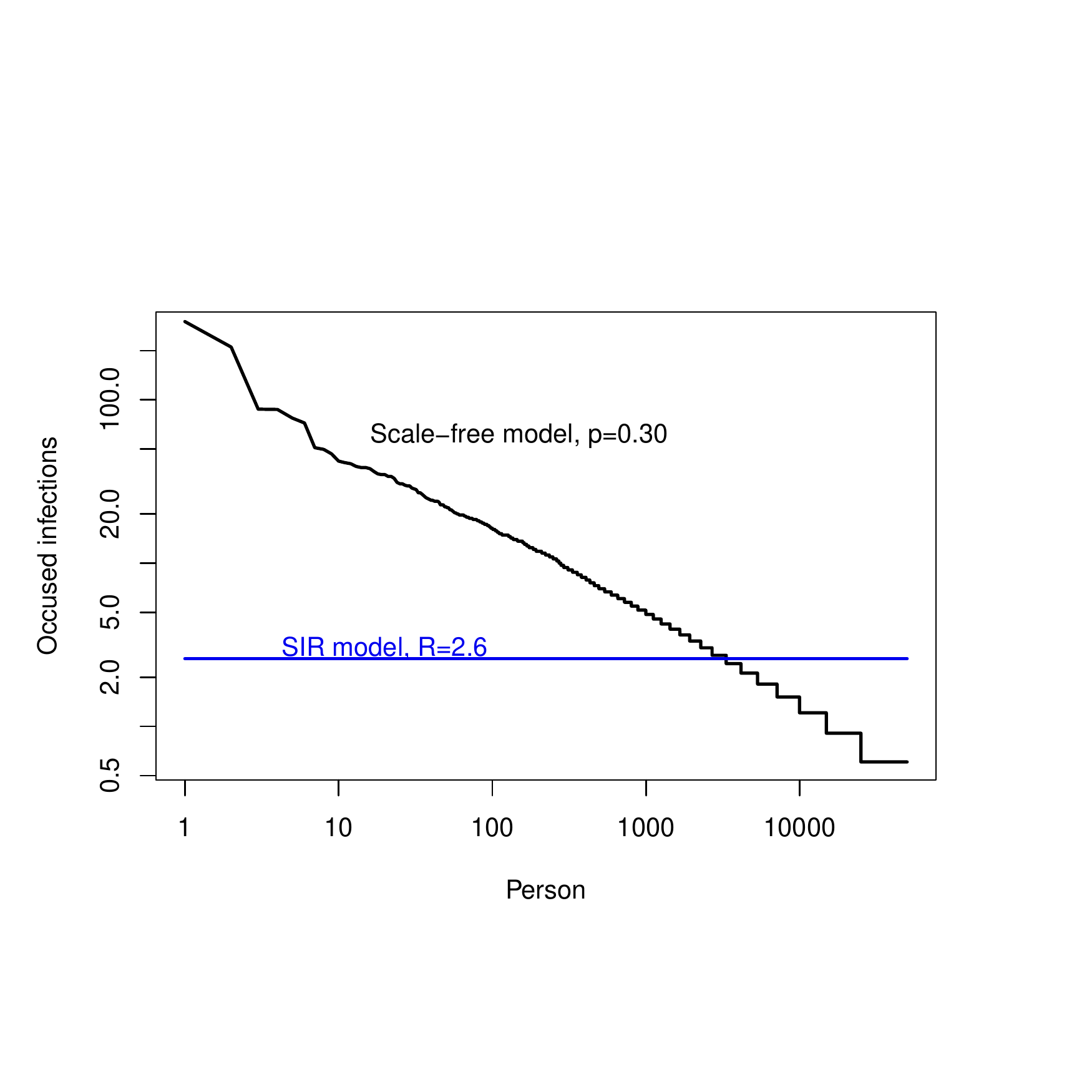}
\caption{Initial distribution of the maximum number of infections a person can cause in the BA and \USIR models. The network behind the black curve is like Fig. \ref{fig:networks} B panel, but for 50,000 persons.}\label{fig:connections}
\end{figure}

The examined community (graph) consisted of 50,000 persons (graph vertices) and their social network (graph edges). The most connected nodes represented the superspreaders.
All existing connections were considered as equally strong and stable in time. The examined scale-free network was a Barab\'asi--Albert network (\cite{2002RvMP...74...47A}) with $m_0=m=2$ parameter selection (BA model in the followings). See Figure \ref{fig:networks} for an illustration, showing similar networks than in our simulations but with 200 points for better perspicuity.

The disease was transmitted via the edges following randomized rules. 
The state of persons could be healthy (who all were susceptible), infected, and recovered (who all had immunity). Ill persons did not transmit the disease to all their connections but only to a fraction of all connections, and this fraction was the same for all persons. Persons with more connections were, therefore, more susceptible to get infection and were more infectious for the community. The propagation rules were:
\begin{itemize}
    \item Healthy persons did not transmit the disease, but become ill if were exposed to an infection. 
    \item Persons who got infection in one step could transmit the infection along each of their connections independently with $p$ probability in the next step. 
    \item{} At the same step when transmitting the infection, ill persons  recovered and got immunity for ever, so they could not get nor transmit the disease.
\end{itemize}

To allow comparisons a SIR mode was also calculated for 50,000 persons. All persons could infect any of the others. The expectation value of $R_0$ was $2.6$ (40\%{} and 60\%{} of the infected population attempted to infect 2 and 3 other persons, respectively).

To get a similar exponential growth in the BA model to the \USIR model, $p=0.30$ was empirically chosen, by simply comparing the upslope of the epidemic curve with various assumptions on $p$ to that of the \USIR model. (Various $p$ values between 0.28--0.32 gave similar quality fits, outside this region the fits were convincingly worse.)

The time unit of a simulation step is the time scale of being infectous, which is in the order of the latency time of a given disease if we assume that people who are already diagnosed with the illness can (and are willing to) behave in a manner that protects other people from being infected. 

After this parameter selection, it was evaluated that persons in the scale-free network could infect initially 1.33 other persons in average, the most active superspreader could infect 331 other persons, the median person infected exactly 1.0 other persons, and the least disease-spreading people could infect 0.67 other people. The distribution of the number of infectable people in the two networks is compared in Fig \ref{fig:connections}

\section{Simulations}
At the beginning of all scenarios 3 persons were infected, chosen randomly from the population. The simulations covered 30 steps. The observed variables were the extension of the epidemic after each step, and the number of new infections that occurred at that step.

In both networks, two scenarios were simulated. The first was a {\it closed} scenario, all new infections were a result of a transmission from an ill person inside the population. The other one was an {\it opened} scenario, where transmission of the disease from outside was possible. In each step during the entire simulation, one randomly chosen person was exposed to infection ``from outside the community'', and become ill if it was susceptible.

\section{Results}

\begin{figure}\centering
\includegraphics[bb=45 150 411 330,width=6.9cm,clip]{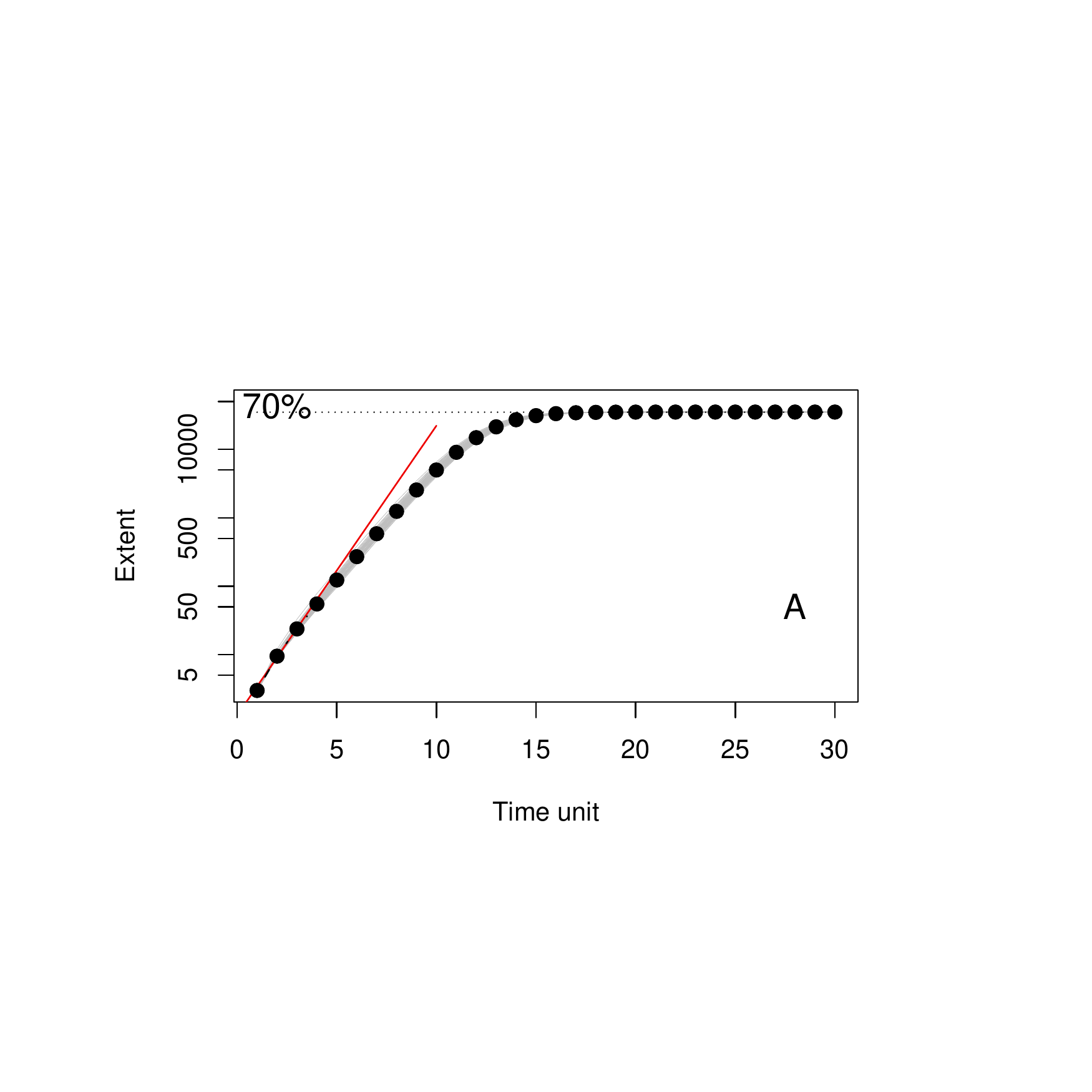}
\includegraphics[bb=45 150 411 330,width=6.9cm,clip]{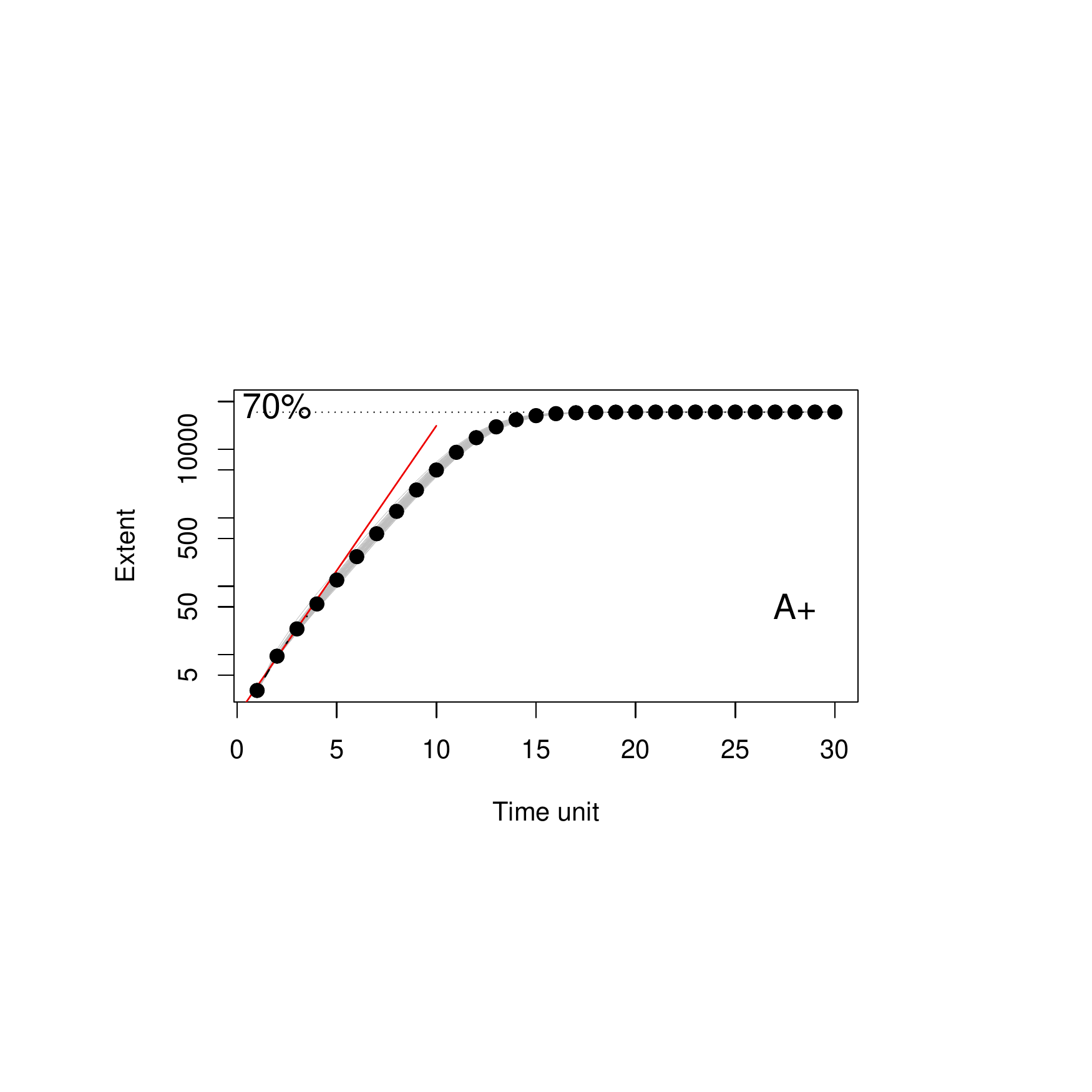}\\
\includegraphics[bb=45 150 411 330,width=6.9cm,clip]{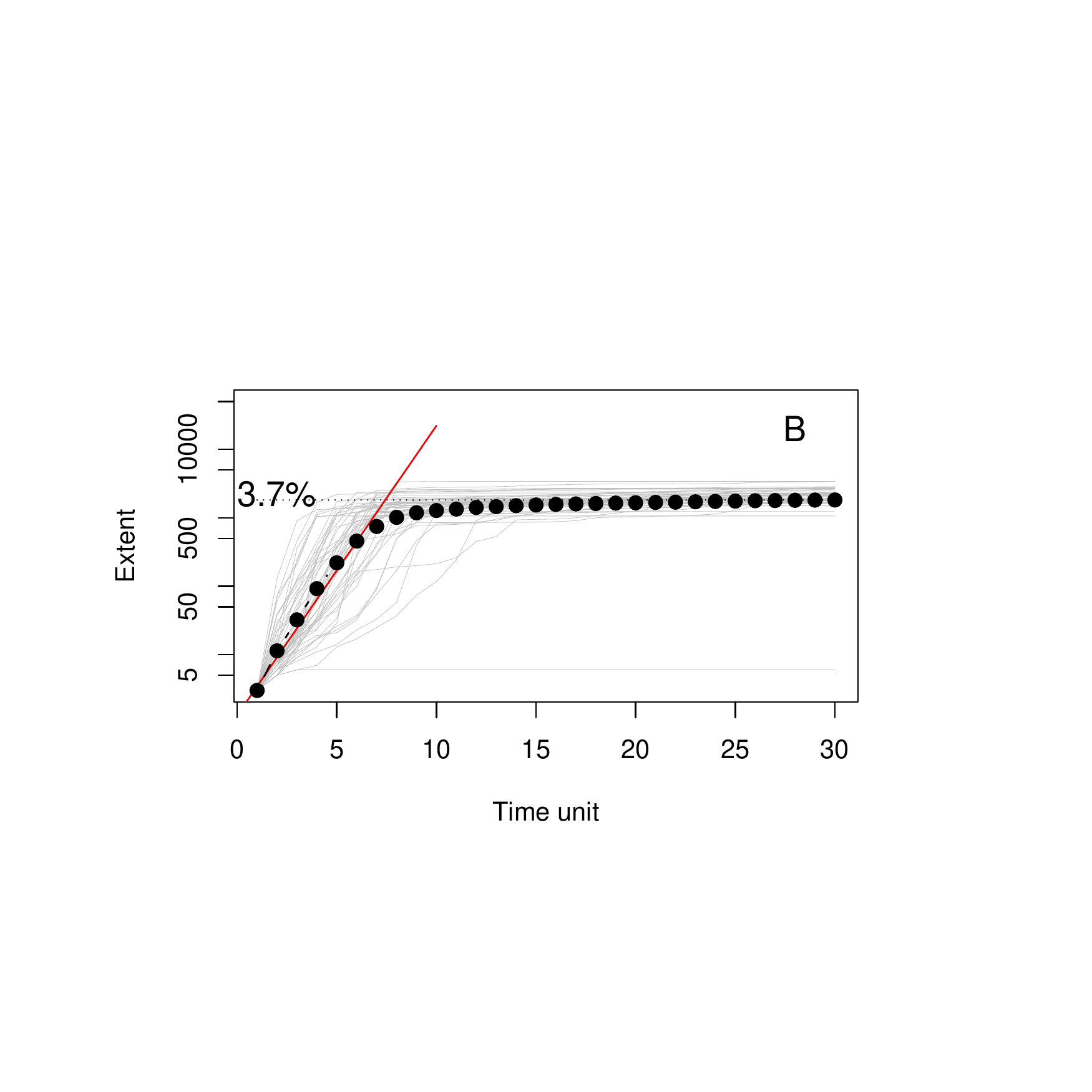}
\includegraphics[bb=45 150 411 330,width=6.9cm,clip]{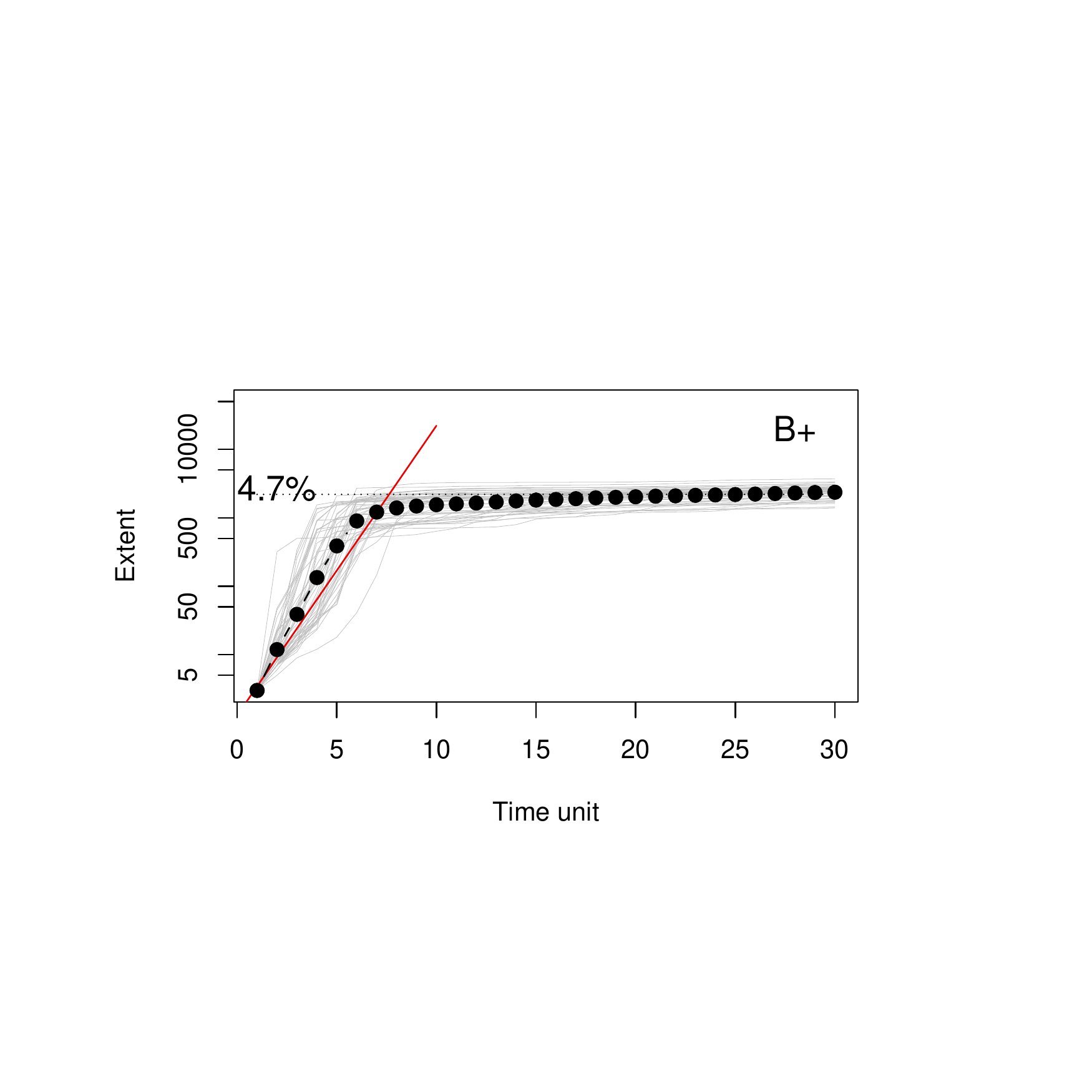}\\
\includegraphics[bb=45 150 411 330,width=6.9cm,clip]{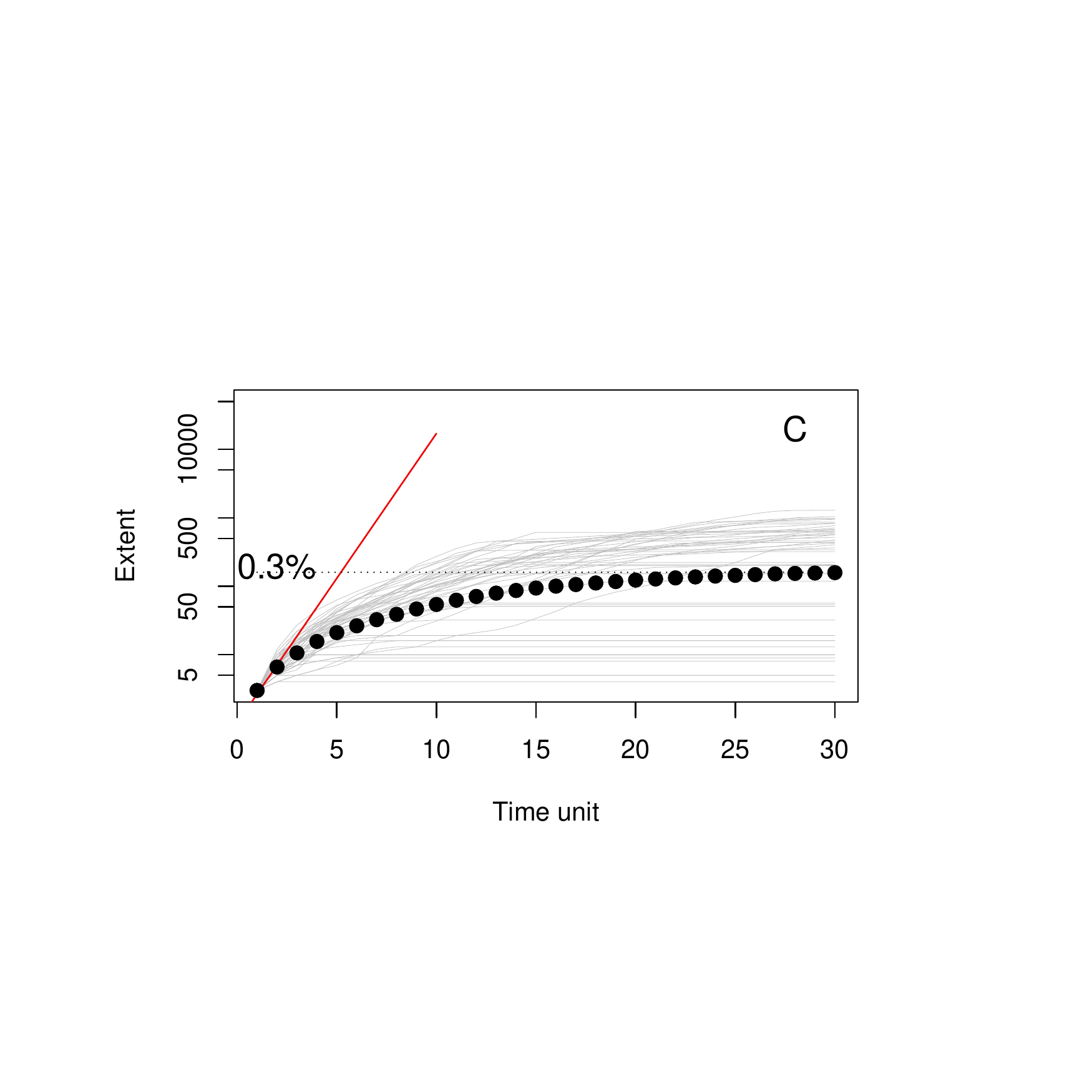}
\includegraphics[bb=45 150 411 330,width=6.9cm,clip]{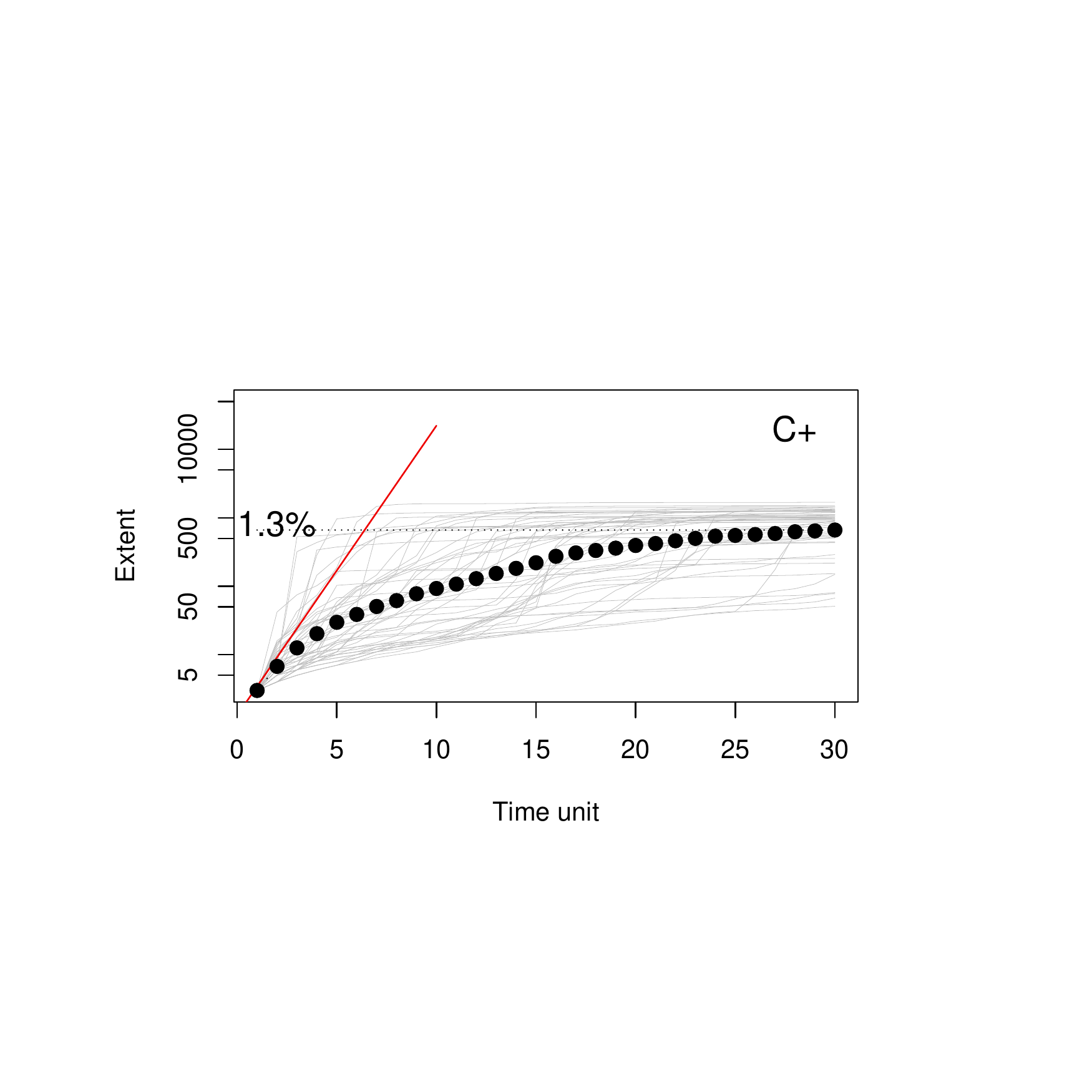}\\
\includegraphics[bb=45 120 411 330,width=6.9cm,clip]{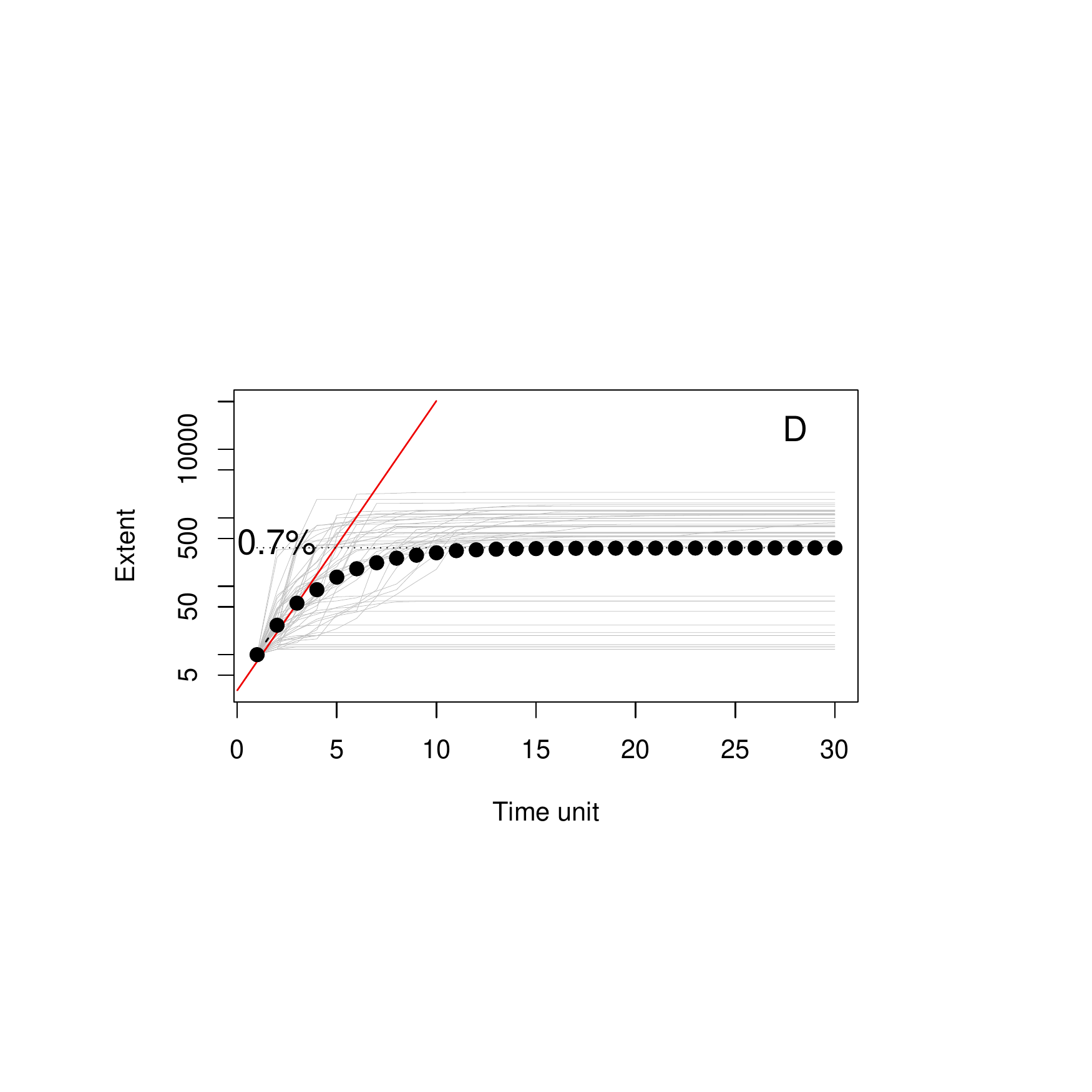}
\includegraphics[bb=45 120 411 330,width=6.9cm,clip]{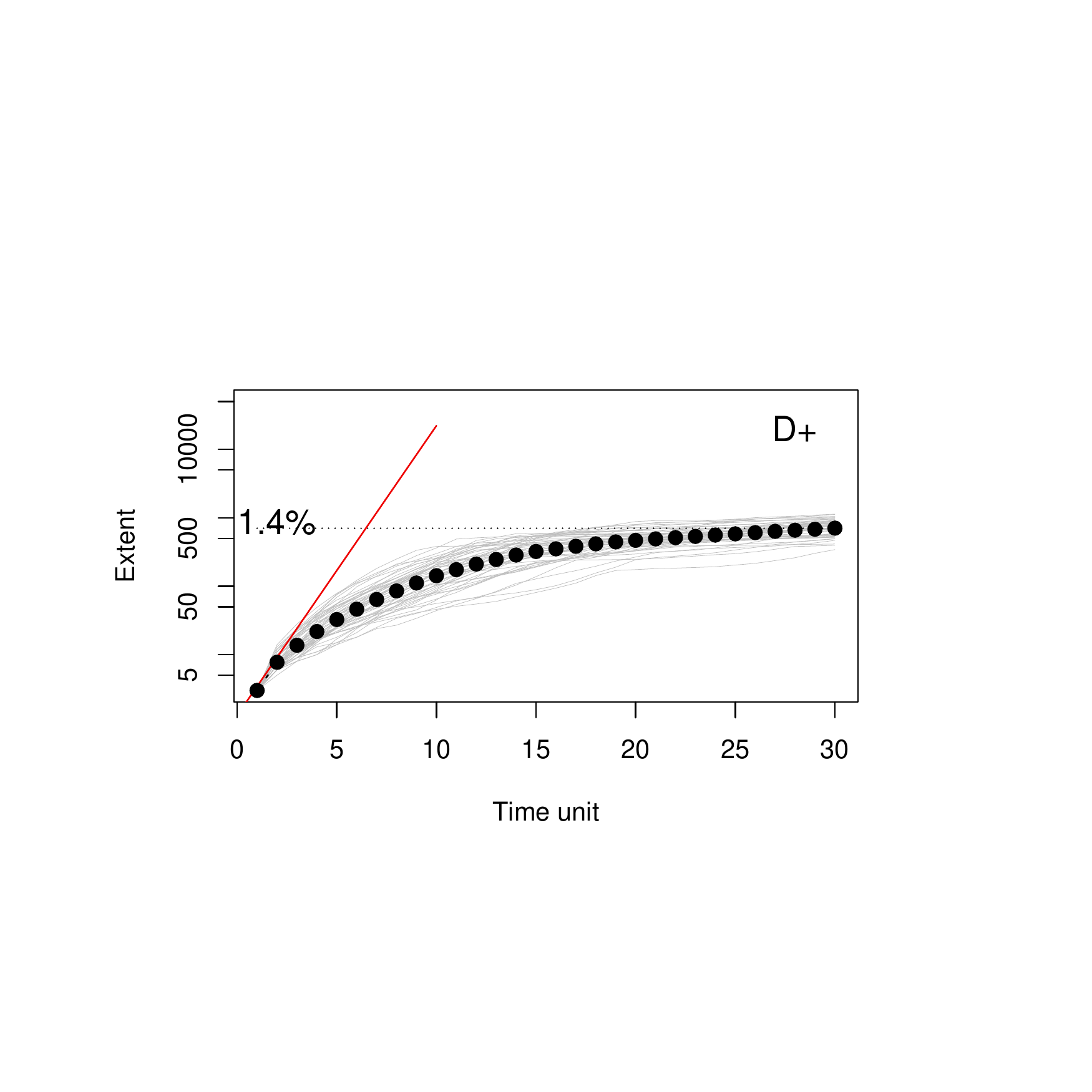}\\
\caption{Cumulative epidemic curves in networks A--D in closed communities (left column) and with disease transportation from outside (right column, labelled with "+" signs). Fifty simulations are plotted in gray, an average scenario at the geometric mean is plotted in black. The red line shows the initial growth rate in the SIR model (A) and the scale-free network without quarantines (B), and with them (C--D). The extent of the epidemic after 30 steps in the different scenarios is indicated inside the left axis. (Note that the vertical axis is logarithmic.)}\label{fig:curves}
\end{figure}

\begin{figure}\centering
\includegraphics[bb=45 150 411 330,width=6.9cm,clip]{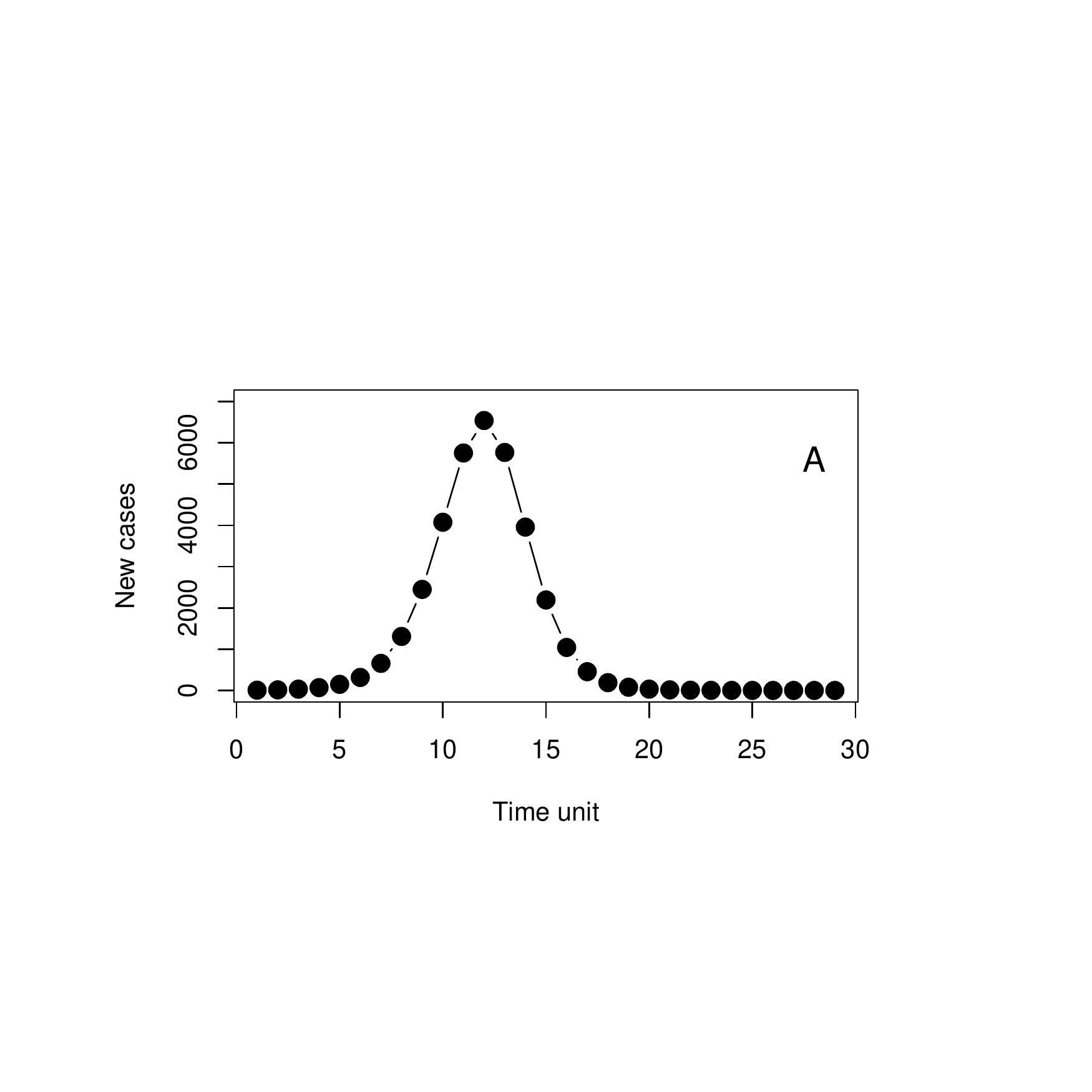}
\includegraphics[bb=45 150 411 330,width=6.9cm,clip]{cases1.pdf}\\
\includegraphics[bb=45 150 411 330,width=6.9cm,clip]{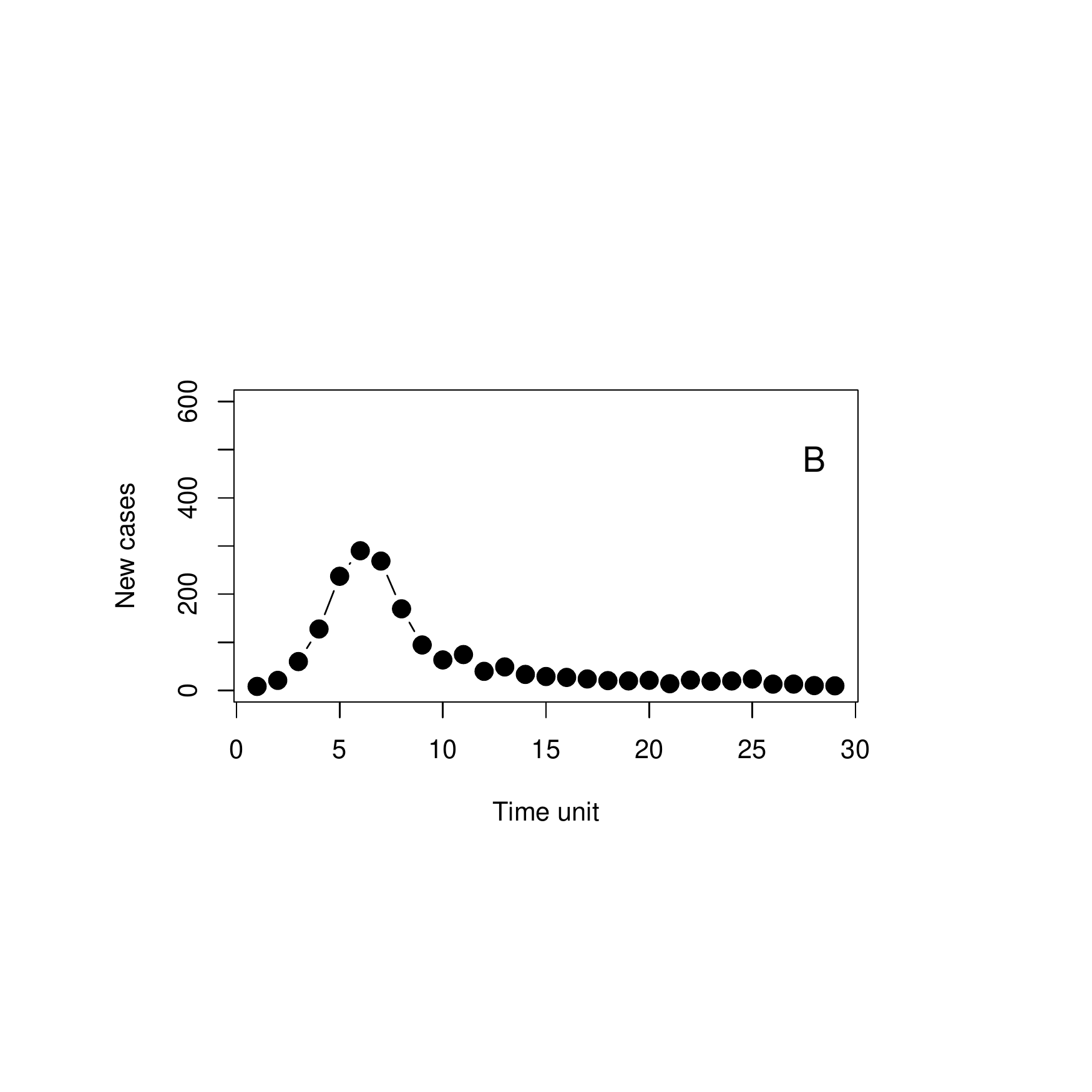}
\includegraphics[bb=45 150 411 330,width=6.9cm,clip]{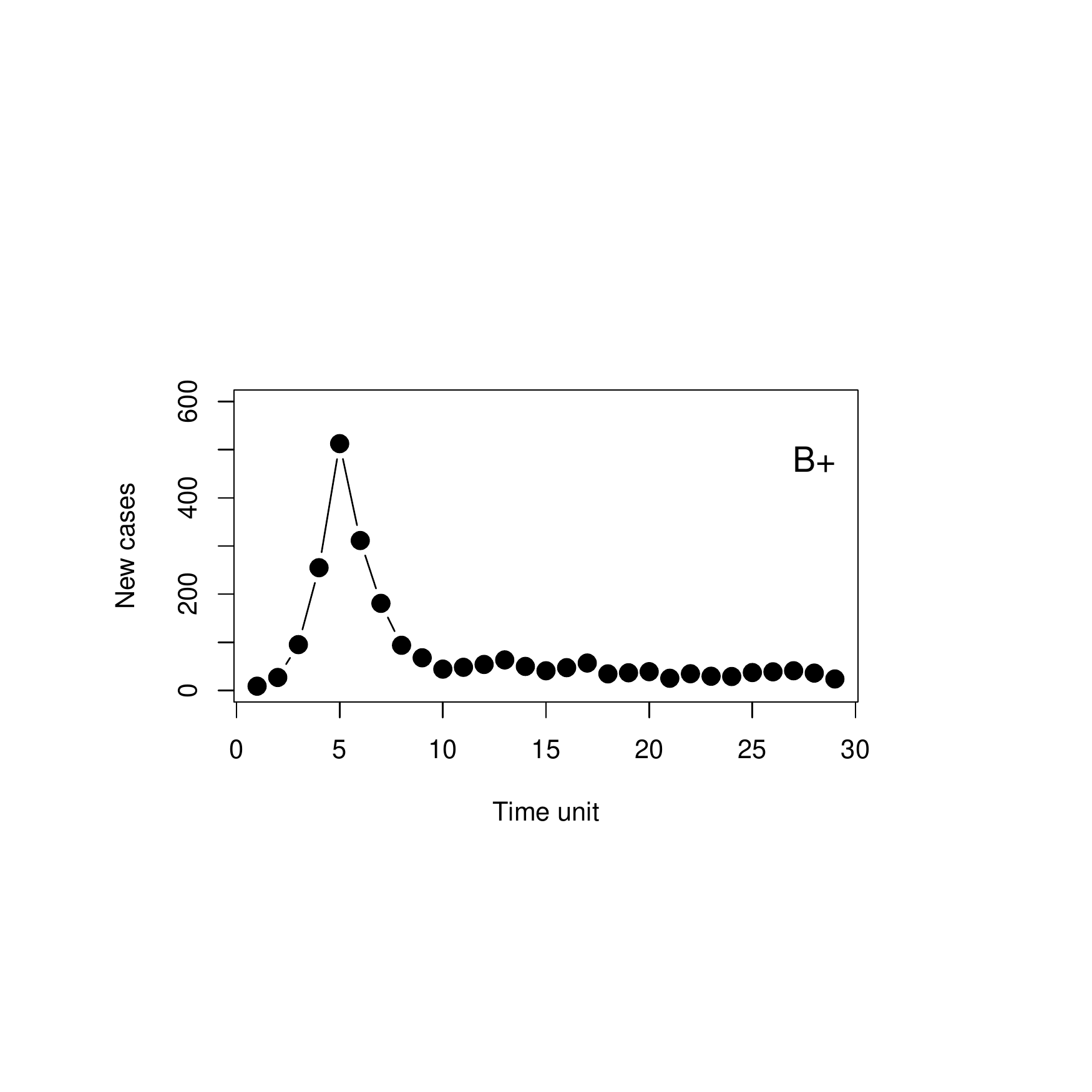}\\
\includegraphics[bb=45 150 411 330,width=6.9cm,clip]{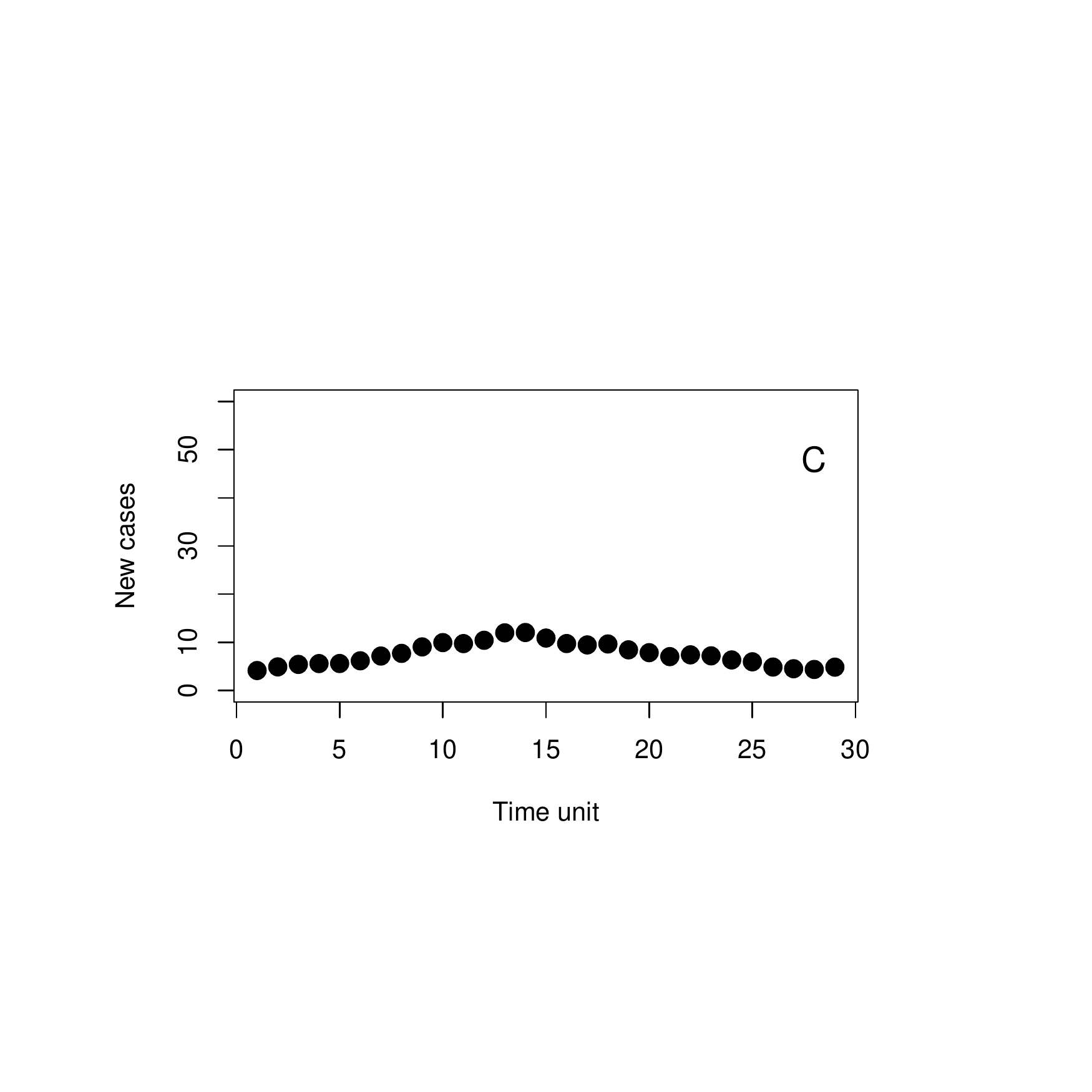}
\includegraphics[bb=45 150 411 330,width=6.9cm,clip]{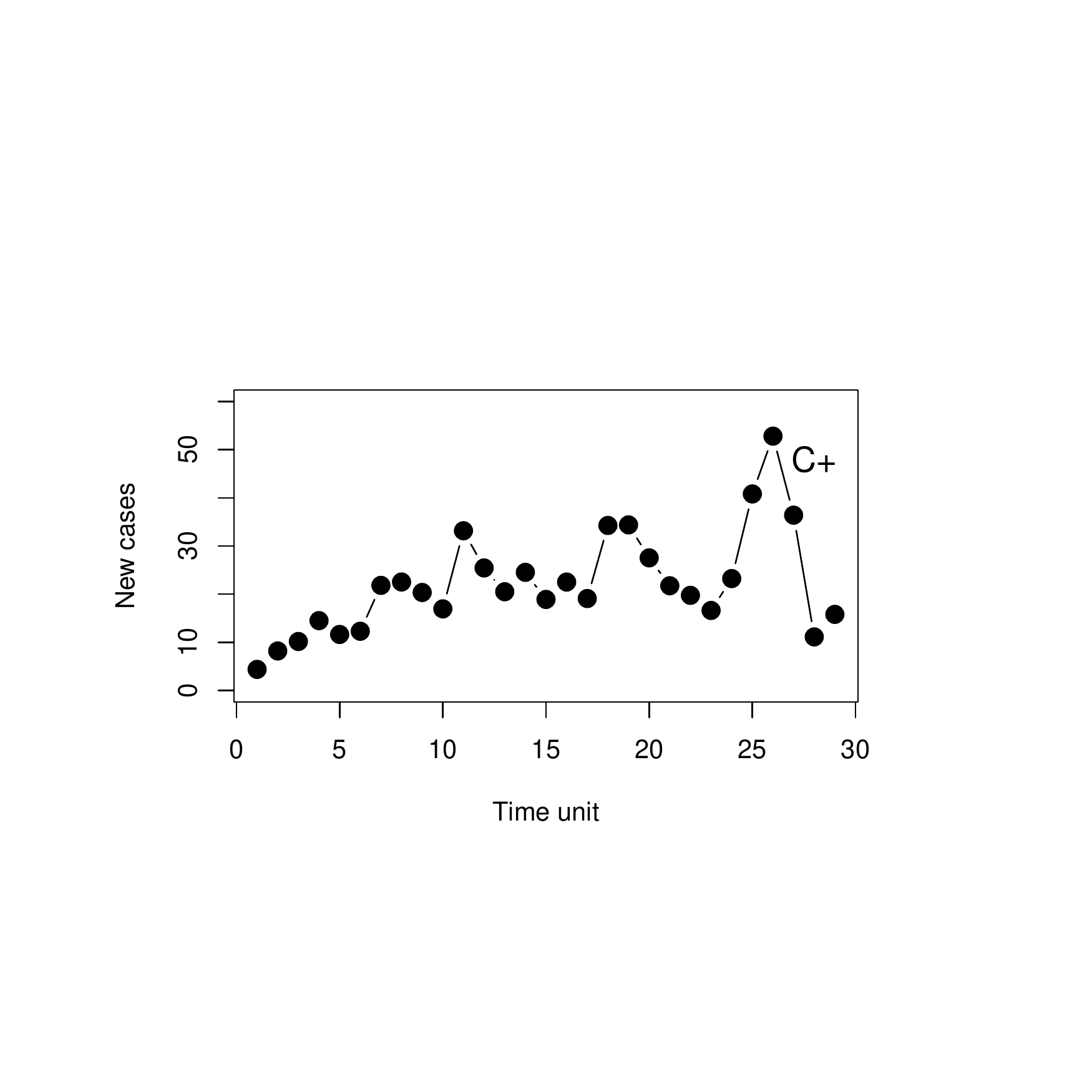}\\
\includegraphics[bb=45 120 411 330,width=6.9cm,clip]{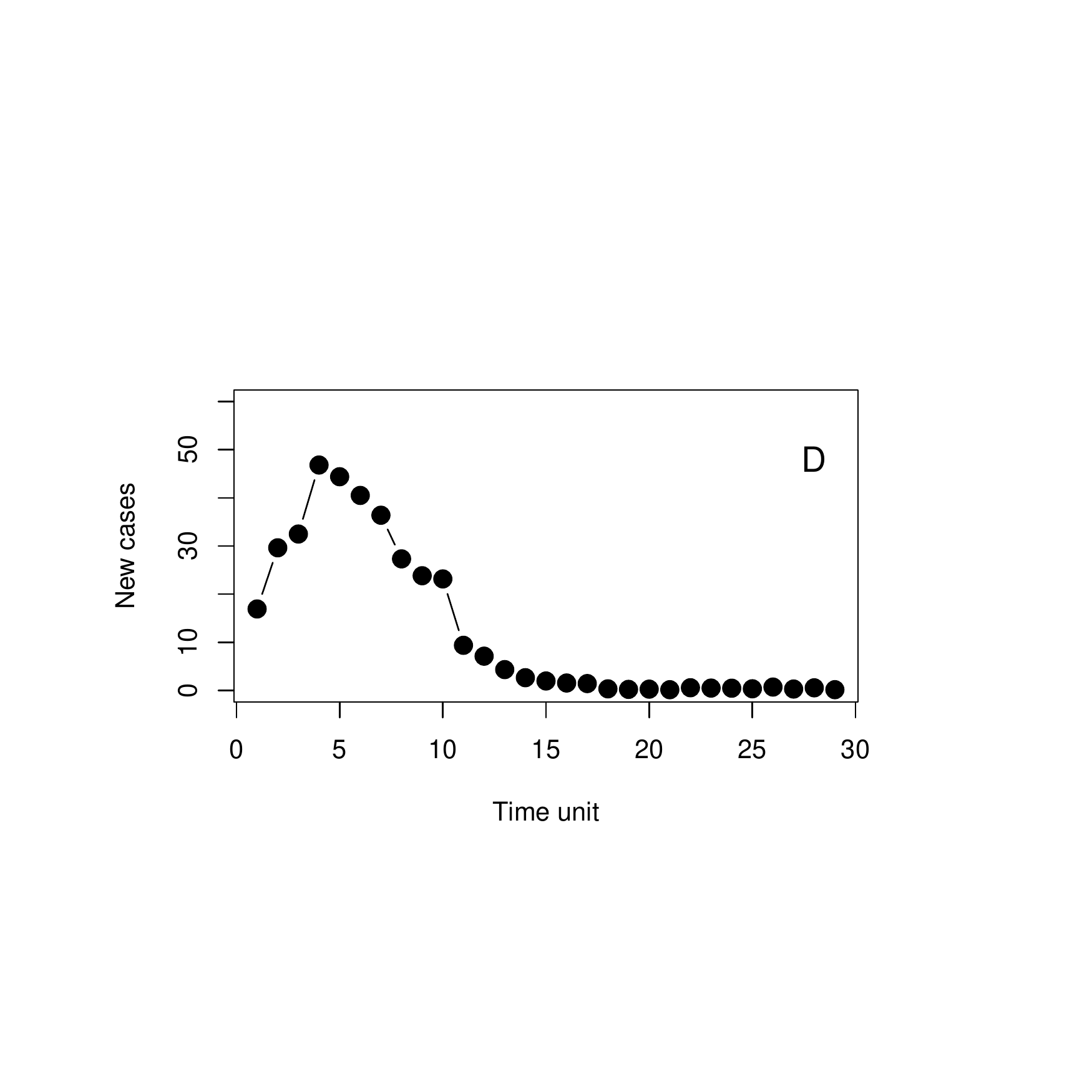}
\includegraphics[bb=45 120 411 330,width=6.9cm,clip]{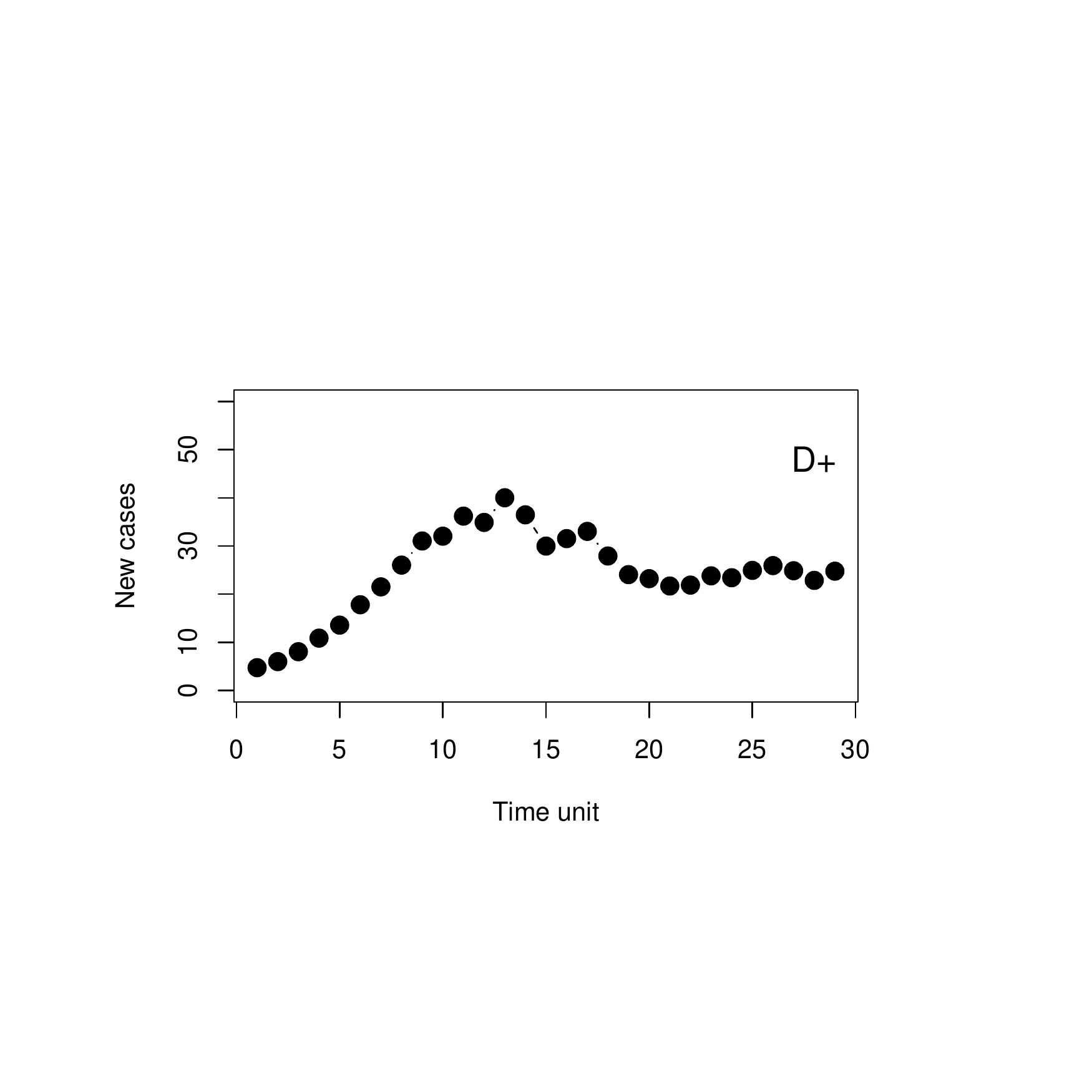}\\
\caption{Epidemic histograms in the SIR model and BA models B--D, in closed communities (left column) and with disease transportation from outside (right column, labelled with "+" signs). (Note that the vertical axis range is the same for the closed and open scenarios, to allow the easier evaluation of transported cases.)}\label{fig:cases}
\end{figure}

The cumulative epidemic curves and histograms are shown in Figures \ref{fig:curves} and \ref{fig:cases}. We can compare the extent of the epidemic after 30 steps, the upslope compared to that of an $R=2.6$ SIR model, the random variations of the epidemic character (50 individual runs are plotted in grey lines) and the mean scenario (plotting the geometric mean of the extent at each steps). We can see that the \USIR simulation in the community reproduces well the SIR model in slope. On the contrary, epidemics in the scale-free network reach a much lesser extent, although the upslope is similarly intensive at the onset. The peak intensity is also at least an order of magnitude lower in the scale-free scenarios than can be prognosticated from the upslope with the SIR model.

\subsection{Accelerated start in the network}

\begin{figure}\centering
\includegraphics[bb=48 110 402 333,width=8cm,clip]{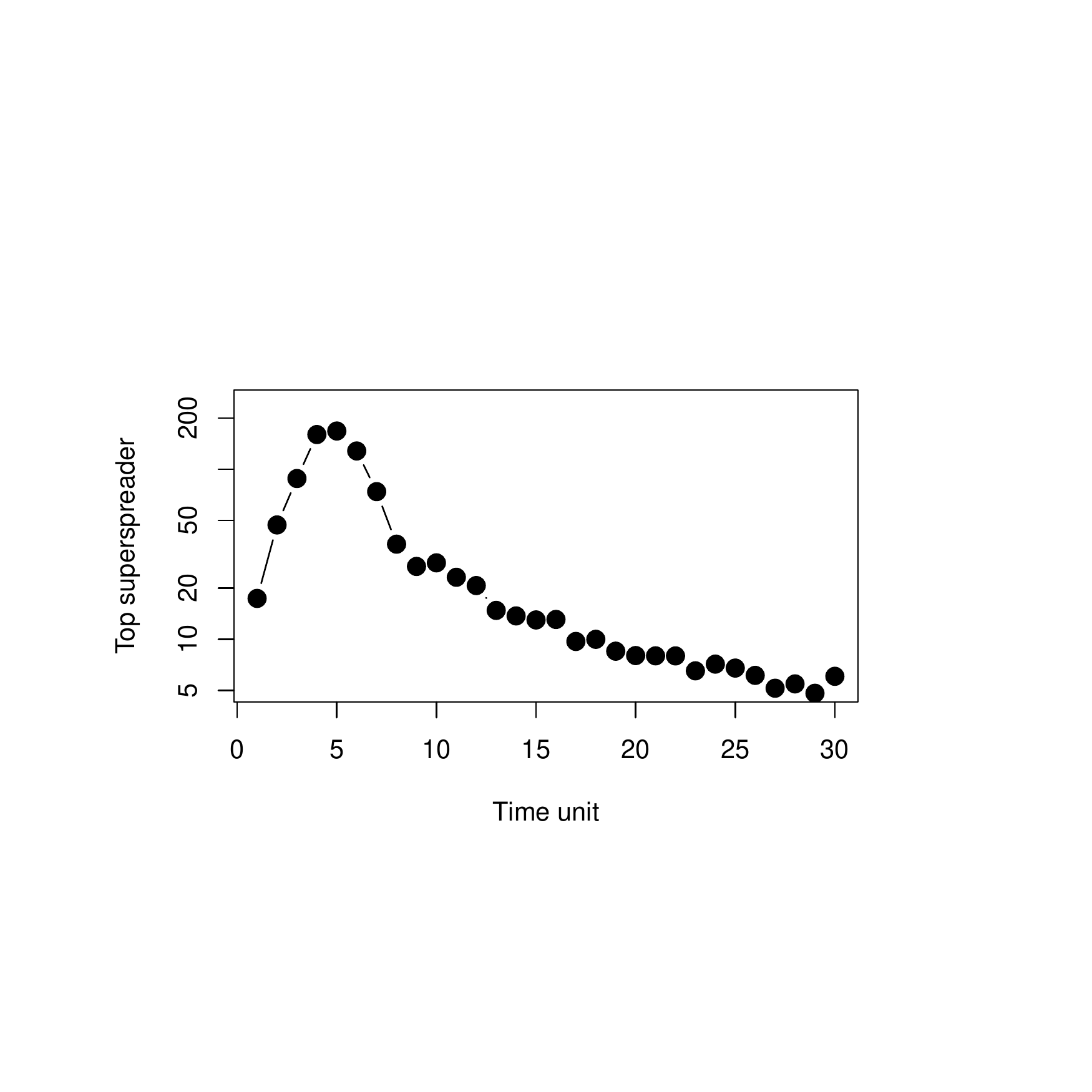}
\caption{Maximum number of connections of the persons infected at the consecutive time units in the B scenario (mean of 200 runs). The top superspreaders get the infection at an early stage.}\label{fig:ssp}
\end{figure}

Although the average number of susceptible persons available for an infected person is lower in a scale-free network than in the \USIR scenario, the epidemic starts equally fast. This is due to the superspreaders in the scale-free network, who have the most connections. After the epidemic starts, the infection trees reach the top superspreaders at an early stage, simply because the points with more connections can be reached earlier. In Figure \ref{fig:ssp} we can see that the maximum number of connections of points infected at the consecutive steps reaches the maximum at an early stage, indicatively after 3--4 time units, and then starts decaying rapidly.  

They can also transmit the disease to many persons, which boosts the upslope of the epidemic curve. This is how the onset of the epidemic in a scale-free model can mimic a larger $R0$ than the expectation number of infections caused by an average person. Therefore, the SIR model applied to predict the extent of epidemic models in a scale-free network leads to an overestimate.

\subsection{Self-mitigation in the network}

Another difference to the SIR models is that the pandemic transmitting in a network has a self-mitigation property. On one hand, this is due to the local pre-immunisation of network. If a person gets infection at a step, it must have come from a connection. When the infected person transmits the disease, that pre-immunised connection will not more be susceptible. This way, the susceptible connections is reduced by at least 1 from the first step.

Also, the development of the epidemics applies an optimal immunisation to the network. \cite{2002PhRvE..65e5103D} proposed that in case of limited immunisation,
the  probability  that  a  certain  node  is  chosen  to  be  immunised  is  proportional  to  its  degree. In all steps in the scale-free models, the susceptible persons get the disease -- and then, immunity -- with a probability that is proportional to the number of their connections, and hence, distributes a fractional immunity in an optimal way. These two mitigation processes explain the prominent differences in the final extent of the epidemic.

\subsection{Mitigation by quarantining}

In scenarios C and D, we can see the effect of quarantining 50\%{} of the nodes blindly, and quarantining only the 1.5\%{} most active superspreaders. Again, left and right panels show closed communities and opened communities with transmitted infections from outside, respectively. These quarantines are effective in slowing down the upslope and the final extent of the epidemic by a factor of 3--6, but it has to be noted that the persons remain in quarantine along the entire simulation, also at the end point. In the open scenarios (C+ and D+) this factor is $\approx$3, which is a factor of 1.5 in the case when half of the community is in quarantine and unavailable for the infection (C+).

\section{Summary and conclusions}

The simulations show that in the cases where the disease propagated in a scale-free network, the epidemic curve followed a different morphology from the SIR model. Initially, the epidemic curve rises steeply in scale-free networks, reflecting that the infection rapidly finds the superspreaders, who transmit the illness to many persons in the population. Therefore, the initially observed exponential increment rate can much exceed the average transmission rate in the population, because the we mostly observe the disease propagation driven by superspreading events. Since $R_0$ is estimated from the realised transmissions at the initial part of the curve, the overrepresentation of superspreading events here overweights the transmission rates of the superspreaders, and heavily biases the estimated $R_0$.

The good news is that most superspreaders get immunity at the beginning of the outbreak, and their transmission rates reduces to zero from then. Therefore, the mean transmission rate in the population, and also the mean transmission rate of the realised infections decrease rapidly, and the exponential rate starts decaying very significantly even without any specific measure to slow down the propagation.

Social distancing has also strong protective effect. Quarantining 50\%{} of the population randomly has similar effects than the succesfull complete isolation of the top 1.5\%{} superspreaders.

It has to be emphasized that these simulations examined only one more-or-less isolated population. If the infection reaches into another closed environment, the evolution in that environment will be similar, and the outcome on the large scale will be the sum of these lower-level scenarios. In this case, multiple outbreaks can also occur. The validity of the results relies on the exact nature of the infection network, which can be different that the one scale-free model I examined here. 


\end{document}